\documentclass[a4paper, amsfonts, amssymb, amsmath, reprint, showkeys, nofootinbib, twoside,superscriptaddress, floatfix,aps,prb,longbibliography]{revtex4-2}
\usepackage[english]{babel}
\usepackage[utf8]{inputenc}
\usepackage[colorinlistoftodos, color=green!40, prependcaption]{todonotes}

\usepackage{amsthm}
\usepackage{mathtools}
\usepackage{xcolor}
\usepackage{graphicx}
\usepackage[left=23mm,right=13mm,top=35mm,columnsep=15pt]{geometry} 
\usepackage{adjustbox}
\usepackage{placeins}
\usepackage[T1]{fontenc}
\usepackage{lipsum}
\usepackage{csquotes}
\usepackage[version=4]{mhchem}
\usepackage{siunitx}  
\usepackage{bm}

\usepackage[pdftex, pdftitle={Article}, pdfauthor={Author}]{hyperref} 
\DeclareSIUnit\formulaunit{f.u.}
\DeclareSIUnit\atom{atom}
\DeclareSIUnit\angstrom{\text {Å}}
\newcommand{\comment}[1]{}
\usepackage{xcolor}
\usepackage{booktabs}
\usepackage{array}

\usepackage{hyperref}
\usepackage[capitalize]{cleveref}
\usepackage{svg}
\crefname{subsection}{subsection}{subsections}
\begin{document}
\title{A neural-network-backed effective harmonic potential study of the ambient pressure phases of hafnia}

\author{Sebastian Bichelmaier}
\affiliation{Institute of Materials Chemistry, TU Wien, A-1060 Vienna, Austria}
\affiliation{KAI GmbH, Europastrasse 8, A-9524 Villach, Austria}
\author{Jesús Carrete}
\affiliation{Institute of Materials Chemistry, TU Wien, A-1060 Vienna, Austria}
\author{Ralf Wanzenb\"ock}
\affiliation{Institute of Materials Chemistry, TU Wien, A-1060 Vienna, Austria}
\author{Florian Buchner}
\affiliation{Institute of Materials Chemistry, TU Wien, A-1060 Vienna, Austria}
\author{Georg K. H. Madsen}
\email[Correspondence email address: ]{georg.madsen@tuwien.ac.at}%
\affiliation{Institute of Materials Chemistry, TU Wien, A-1060 Vienna, Austria}

\date{\today} 

\begin{abstract}
Phonon-based approaches and molecular dynamics are widely established methods for gaining access to a temperature-dependent description of material properties. However, when a compound's phase space is vast, density-functional-theory-backed studies quickly reach prohibitive levels of computational expense. Here, we explore the complex phase structure of \ce{HfO2} using effective harmonic potentials based on a neural-network force field (NNFF) as a surrogate model. We detail the data acquisition and training strategy that enable the NNFF to provide almost ab-initio accuracy at a significantly reduced cost and present a recipe for automation. We demonstrate how the NNFF can generalize beyond its training data and that it is transferable between several phases of hafnia. We find that the thermal expansion of the low-symmetry phases agrees well with experimental results and we determine the $P\bar{4}3m$ phase to be the favorable (stoichiometric) cubic phase over the established $Fm\bar{3}m$. In contrast, the experimental lattice constants of the cubic phases are substantially larger than what is calculated for the corresponding stoichiometric phases. Furthermore, we show that the stoichiometric cubic phases are unlikely to be thermodynamically stable compared to the tetragonal and monoclinic phases, and hypothesize that they only exist in defect-stabilized forms. 
\end{abstract}


\maketitle

\section{Introduction} \label{sec:introduction}
The intricate link between microscopic crystal structures and macroscopic materials properties necessitates understanding the former to predict the latter. Temperature is an important contributor to phase stability. However, computational studies of its impact generally require extensive sampling and with that a large number of energy and force evaluations. Hence, temperature dependence has been notoriously difficult to combine with accurate computational methods like density functional theory (DFT) \cite{Glensk_PRX14}.

Effective harmonic potentials (EHPs) are a widely used method to include temperature dependency into ab-initio studies \cite{errea_PRL_2013,hellman_PRB_2013,Stern_PRB16,roekeghem_prx_16,Monacelli_JPCM21,roekeghem_CPC_2021,bichelmaier_22}. An EHP is constructed by sampling the potential energy surface (PES) using temperature-dependent displacement distribution functions and fitting effective force constants. Performing the described procedure with a DFT backend is computationally demanding, especially for low-symmetry structures. Thus, while the method gives access to structure- and pressure-dependent free energies, its application to the calculation of phase transitions has been somewhat limited \cite{Monacelli_JPCM21}. It would be an obvious advantage if the sampling of the PES inherent to the EHP could be used to simultaneously train a surrogate model.

Recently, machine-learning (ML) force fields (FFs) are becoming increasingly widespread in the field of computational materials science \cite{unke_21,ko21,goedecker1}, 
with applications ranging from an accurate description of the martensitic phase transition in the memory alloy \ce{NiTi} \cite{tang_22_niti} or grain boundaries in copper systems \cite{yunzhen_22_cu} to structural phase transitions \cite{PhysRevB.105.064104} and surface reconstructions \cite{wanzenboeck} in \ce{SrTiO3}. Given sufficient training, MLFFs can replace DFT calculations at a vastly reduced cost, and combining them with EHPs can lead to significantly lower computational demands. 

Hafnia (\ce{HfO2}) has recently attracted particular research interest \cite{huan_PRB_2014,batra_16,batra_jpc_2017,tobase_pss_18,ding_20,behara_22,bichelmaier_22}. Similar to the isostructural material zirconia (\ce{ZrO2}) it boasts a rich phase diagram with various internal (doping, alloying) and external (strain, temperature) factors contributing to phase stability and it is thus an ideal use case for MLFFs \cite{wu_prb_21,sivaraman_npj_20}. At ambient pressure and temperatures up to approximately \SI{2050}{\kelvin} the material presents a monoclinic ($P2_1c$) crystal structure (m-phase), which, as temperature increases, undergoes a first-order transition into a tetragonal ($P4_2nmc$) one (t-phase) \cite{massalski1986binary,wang_jamcersoc_06}. 
Furthermore, hafnia is believed to transition to a high-symmetry cubic ($Fm\bar{3}m$) phase (cI-phase) in a narrow temperature window immediately below the melting point \cite{fan_jpc_19}.

There is, however, still quite some ambiguity regarding the transition to a high-symmetry cubic phase. The reported t-to-c phase transition temperatures for \ce{HfO2} span almost \SI{1000}{\kelvin} \cite{wang_jamcersoc_06}, and there is even some doubt regarding the existence of a pure cubic phase or its precise space group \cite{kuenneth_pss}. Indeed, some researchers find the lower-symmetry $P\bar{4}3m$ (cII-phase) phase to be energetically favorable \cite{Barabash2017}, although to our knowledge no theoretical studies, that properly take temperature into account, exist. For \ce{HfO2}, few experimental studies describing a t-to-c phase transition exist, potentially due to the extreme temperature requirements and the difficulties these conditions present for accurate experimental analysis. Nonetheless, some researchers find a mixed phase of cubic and tetragonal symmetry suggesting metastability \cite{sivaraman_prl_21}, while others see a clear second-order transition \cite{tobase_pss_18}. It should be pointed out that similar ambiguities have been reported with respect to the existence of a stable stoichiometric cubic phase in the related \ce{ZrO2} \cite{mcclellan_pma_94,wang2004,sternik_jcp_05,ohtaka_jac_05,tolborg}. Thus, additional theoretical efforts are warranted, not only to clarify if a stable stoichiometric cubic phase exists, but also to shed light on its crystal structure.

In this work, we build an MLFF using a neural network (NNFF) with a methodology corresponding to an evolved version of the one presented in Ref.~\onlinecite{carrete_jcim21}. We construct a NNFF that can describe the part of the potential energy landscape studied well, and then employ it as the backend calculator for an EHP-based treatment of \ce{HfO2}. We analyze the m-to-t phase transition and establish a case for $P\bar{4}3m$ being the more likely candidate for a high-T cubic phase in \ce{HfO2} over the traditionally assumed $Fm\bar{3}m$. The full thermodynamical picture however suggests that neither of the cubic phases is completely stable in purely stoichiometric conditions, but may be stabilized by oxygen vacancies.

We furthermore discuss EHPs for training set generation. Generation of a proper training set for MLFFs is far from trivial \cite{Zaverkin_DD22}. Molecular dynamics (MD) trajectories typically consist of strongly correlated structures thus subsampling very long trajectories is required, resulting in hundreds of DFT calculations not used in training. Some authors combine this with active learning, using a subsampled trajectory as a baseline and an uncertainty metric, such as a committee of networks, to guide additional calculations \cite{deepmd_2,Jinnouchi_JPCL20,Schwalbe-Koda2021}. Continuously retraining NNFFs with often only marginal augmentations to the data is computationally demanding and the accuracy of committee-based error estimates is still under debate \cite{scalia_uncertainty_2020}.
Here, we show a training strategy based on EHPs, whereby the inherent physically meaningful and iterative sampling of the PES can be exploited. Thereby an NNFF is trained systematically on the cumulative data obtained up to each iteration, mimicking an active learning approach.

\section{Method}\label{sec:method}
\subsection{Effective harmonic potentials}
As presented in detail elsewhere \cite{bichelmaier_22}, an effective harmonic potential $\mathcal{V}$ and its associated density matrix $\hat{\rho}$ can be obtained through a variational approach \cite{errea_PRL_2013,Errea_PRB14,Monacelli_JPCM21}, resulting in a parametrization of the optimal harmonic potential in terms of the second-order force constants, $\mathbf{\Phi}$:
\begin{equation}
    \mathcal{V}(\mathbf{u}) =  \frac{1}{2} \mathbf{u}^\mathrm{T} \mathbf{\Phi} \mathbf{u}
    \label{eqn:v}
\end{equation}
out of which a corresponding density matrix

\begin{equation}
    \rho(\mathbf{u}) = \frac{1}{\sqrt{(2\pi)^{3N}\left|\mathbf{C}\right|}} \exp\left(-\frac{1}{2}\mathbf{u}\mathbf{C}^{-1}\mathbf{u}\right),
    \label{eqn:dm}
\end{equation}
can be built. Here $\mathbf{u}$ refers to the mass-weighted displacements of the atoms, $\mathbf{u}=\sqrt{M_i}\left(\mathbf{R}_i-\mathbf{R}_{i,0}\right)$. The elements of the covariance matrix, $C_{ij}$ are obtained as

\begin{equation}
    C_{ij} = \frac{\hbar}{2\sqrt{M_iM_j}}\sum_\lambda \frac{1}{\omega_\lambda \tanh{\frac{\hbar \omega_\lambda}{2k_BT}}}\epsilon_{\lambda, i}\otimes \epsilon_{\lambda, j}^*,
    \label{eqn:covmatrix}
\end{equation}
where $\omega_\lambda$ and $\bm{\epsilon}_\lambda$ are the eigenvalues and -vectors of the second-order force constants [\cref{eqn:v}] and $M_i$ is the mass of atom $i$.
The free energy, $\mathcal{F} = F_{\text{harm}} + F_{\text{corr}}$, consisting of a harmonic contribution and an anharmonic correction term, can then be expressed as
\begin{gather}
    F_\text{harm}(T) = \sum_\lambda \left(\frac{\hbar \omega_\lambda}{2} + k_{\text{B}} T \log\left[1-\exp\left\{-\frac{\hbar \omega_\lambda}{k_{\text{B}}T}\right\}\right]\right)\label{eqn:harmonic:free}\\
    F_{\text{corr}}(T) = \frac{1}{N}\sum_{n}  \left[V(\mathbf{u}_n) - \mathcal{V}(\mathbf{u}_n)\right] ,
    \label{eqn:correction}
\end{gather}
 with $\mathbf{u}_n$ corresponding to samples drawn from the real-space distribution defined by the density matrix, \cref{eqn:dm}, and $V$ to the true energy as obtained from DFT.

\subsection{Neural-network force field}
\subsubsection{Architecture and loss}
 We use a NeuralIL architecture \cite{carrete_jcim21}, with a cutoff-radius of \SI{5}{\angstrom}, an embedding dimension of $4$ and a total of $28$ basis functions in a perceptron consisting of $128:64:32:16:16:16$ neurons. We employ a log-cosh based loss \cite{Wang2022} for the forces
 \begin{equation}
    \begin{split}
     \mathcal{L}_f =\frac{0.1\,\text{eV}\text{\AA}^{-1}}{n_\text{atoms}}\sum_i^{n_\text{atoms}}\! \log\left[\cosh\left(\frac{\sqrt{\frac{1}{3}\sum_\alpha \Delta f_{i,\alpha}^2}}{0.1\,\text{eV}\text{\AA}^{-1}}\right)\right],
     \end{split}
     \label{logcosh_f}
 \end{equation}
 where $\Delta f$ is the difference between the actual force as obtained from DFT and the NNFF-predicted force, $\alpha$ indexes the three Cartesian axes and $i$ the atoms.
 
 While the forces in principle contain enough information to model the potential energy surface, including an energy loss can enable more efficient learning of multiple local minima. Furthermore, once unit cell volume is included as a variable, a continuum of structures exhibiting zero forces, but with different energies, needs to be accurately described. Hence, an energy loss can be included in the overall loss
  \begin{equation}
    \begin{split}
     \mathcal{L}_E =\frac{0.1\,\text{eV}}{{n_{\text{atoms}}}}\log\left[\cosh\left(\frac{\Delta E}{0.1\,\text{eV}}\right)\right],
     \end{split}
     \label{logcosh_E}
 \end{equation}
 where $\Delta E$ corresponds to the error of the predicted energy as compared to the DFT result.
 The obvious approach of combining \cref{logcosh_E,logcosh_f} is to introduce a new hyperparameter, $w_E$, tuning the fraction of the energy loss included in the overall loss:
 \begin{equation}
     \mathcal{L}_{\mathrm{tot,manual}} = \left(1-w_E\right)\mathcal{L}_f + w_E\mathcal{L}_E.
     \label{eqn:loss_manual}
 \end{equation}
 However, as the loss measures the quality of the NNs approximation of the PES, the optimal choice of weight might change or depend on the composition or the dataset of the compound studied. Furthermore, manually tuning a hyperparameter is time- and resource-intensive. 
 
 Several approaches have been developed to overcome these obstacles. We implemented and tested the inverse certainty weighting \cite{wang_2021} and the inverse Dirichlet weighting \cite{maddu_22}, but both resulted in instabilities during training, which manifested themselves as very large energy weights and essentially halted any progress. The most stable and efficient approach tested for the present datasets was Mitrevski's \cite{mitrevski_moo_2020} implementation of the stochastic multi-gradient descent algorithm \cite{liu_2019}. Here the gradients of \cref{logcosh_E,logcosh_f}, $\nabla\mathcal{L}_f$ and $\nabla\mathcal{L}_E$, are corrected using the first and second momenta (\textit{adamizing}, \cite{mitrevski_moo_2020}), resulting in smoothed contributions of the force ($\tilde{\mathcal{L}}_f$) and energy loss ($\tilde{\mathcal{L}}_E$). Subsequently the common descent vector is calculated 
 \begin{equation}
     \nabla\mathcal{L} = \left(1-\omega_E\right) \nabla\tilde{\mathcal{L}}_f + \omega_E \nabla\tilde{\mathcal{L}}_E
     \label{eq:moo}
 \end{equation}
 where $\omega_E$ is obtained through a quadratic optimization
 \begin{equation}
     \min\limits_{\omega_E}\left[\left\lVert \omega_E \nabla \tilde{\mathcal{L}}_E + \left(1-\omega_E\right)\nabla\tilde{\mathcal{L}}_f \right\rVert^2\quad  0\leq\omega_E\leq 1\right].
     \label{eq:weights}
 \end{equation}
 
The use of a common descent vector has been shown to result in a stable and efficient multi-objective optimization and enables exploration of the Pareto border \cite{mitrevski_moo_2020}.
To validate the method in the context of our study, we trained a neural network and stored the weights over all epochs and batches. The resulting NN is compared to two NNs trained with the average of these gathered weights $\omega_E \approx 0.273$, one using just the adamized loss, \cref{eq:moo}, and ignoring \cref{eq:weights} and one using \cref{eqn:loss_manual}. Indeed, training NNs with the fixed energy weight results in a parametrization with similar performance 
as applying \cref{eq:moo,eq:weights}. 
The advantage thus lies in the elimination of the $\omega_E$ hyperparameter, which otherwise would have to be manually tuned. 

The loss described above is minimized over a total of $1000$ epochs using a one-cycle scheduler, varying the learning rate linearly from $1.5\times10^{-4}$ to $1.5\times10^{-3}$ and back, with the final \SI{10}{\percent} decreasing further to $1.5\times10^{-5}$.

\subsubsection{Augmentation}
Due to the high dimensionality of the sampling space described earlier, we find forces in the training set ranging from a few \SI{}{\milli\electronvolt\per\angstrom} to several hundreds of \SI{}{\electronvolt\per\angstrom}, caused by atoms being close to each other. As it is impossible and undesirable to enumerate a sufficient number of atomic environments resulting in those high forces and the NNFF behaving potentially erratically in untrained areas we include a repulsive contribution as suggested in \cite{bartok_2018}. We choose the repulsive part of the Morse potential to provide some robustness in those sparsely sampled regions:
\begin{equation}
    V_{\text{rep}}(\mathbf{r}) = \frac{1}{2}\sum_{i\neq j} d_{ij} \times h( r_{ij}, r_s, r_c)e^{-2a_{ij}\left(r_{ij}-b_{ij}\right)},
    \label{eqn:morse}
\end{equation}
where $r_{ij}$ is the distance between atoms $i$ and $j$, $r_s$ is a trainable switching radius and $r_c$ the cutoff radius, set equal to the cutoff radius of the NNFF, i.e. $r_c=\SI{5}{\angstrom}$. The $a_{ij}, b_{ij}, d_{ij}$ represent element-specific parameters which are optimized during the training process if atom $i$ and $j$ are of the same type. For interactions between two different chemical species, the corresponding ``mixed'' parameters obtained through the rules established in Ref.~\onlinecite{kong73} are used. The bump function, $h$, is defined by 
\begin{gather}
  f(x) = \begin{cases}
    0 & x \leq 0 \\
    \exp\left(-\frac{1}{x}\right) & x > 0
  \end{cases}\\
  g(x) = \frac{f(x)}{f(x) + f(1-x)} \\
  h(r, r_s, r_c) = 1 - g\left(\frac{r^2 - r_s^2}{r_c^2-r_s^2}\right) .
  \label{eq:bump}
\end{gather}

\begin{figure}
    \centering
    \includegraphics[width=\columnwidth]{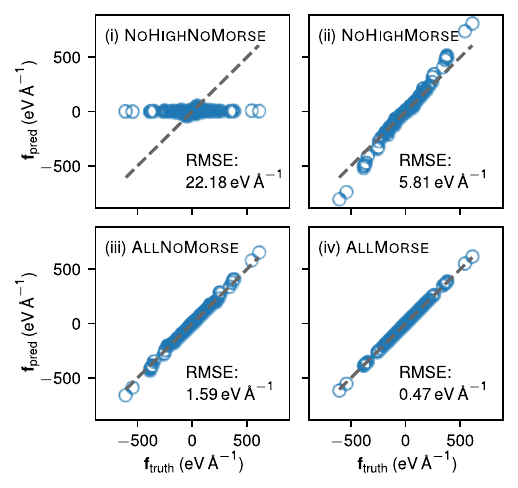}
    \caption{Comparison of the NNFF with and without the Morse model evaluated on high-force configurations, with training data including and excluding high-force configurations.}
    \label{fig:nomorse_morse}
\end{figure}

To test how efficiently a NNFF can be built to be useful as a surrogate model for EHPs, we decided to analyse a subset of the DFT data generated for our investigation of cI-\ce{HfO_2} from \cite{bichelmaier_22}. As the cubic phase shows an instability at \SI{0}{\kelvin}, the first iteration data contains numerous high-energy configurations with large forces. We use a subset of the first $5$ iterations, resulting in $172$ structures, to validate this approach and train four NNFFs: using (i) only those configurations where the norms of all forces are below \SI{50}{\electronvolt\per\angstrom} and without \cref{eqn:morse} (\textsc{NoHighNoMorse}), (ii) only those configurations where the norms of all forces are below \SI{50}{\electronvolt\per\angstrom} and with \cref{eqn:morse} (\textsc{NoHighMorse}), (iii) all configurations without Morse (\textsc{AllNoMorse}) and (iv) all configurations with Morse (\textsc{AllMorse}). 
The resulting parity plots on the high-force configurations are shown in \cref{fig:nomorse_morse}. As expected, the \textsc{NoHighNoMorse} NNFF performs poorly; in fact, it predicts a negligible force where large forces should be expected. This would lead to quickly-failing sampling as an artifactual minimum state with clusters of atoms would be considered advantageous. The \textsc{NoHighMorse} potential, on the other hand, predicts at least \emph{some} high force. As the exact value in this case is not necessarily important the addition of the term \cref{eqn:morse} can indeed be understood as fulfilling its intended purpose: providing robustness in sparsely sampled regions. Naturally, for production runs we include all data, where the addition of the Morse potential proves superior as well.
 
\subsection{Computational details}
The DFT data used throughout this manuscript is obtained using the projector-augmented-wave (PAW) formalism \cite{bloechl_PRB_1994} as implemented in VASP \cite{kresse_PRB_1996} using the PBE functional \cite{pbe1}, an energy cutoff of \SI{600}{\electronvolt} and the 5p6s5d and 2s2p as valence states for \ce{Hf} and \ce{O} respectively. Supercells of size $4\times4\times4$, $4\times4\times3$, $2\times2\times2$ and $2\times2\times2$ were used for the $Fm\bar{3}m$, $P4_2nmc$, $P\bar{4}3m$ and $P2_1c$ phases, respectively. A $\Gamma$-only $k$-point mesh was used for all calculations. Doubling the $k$-point mesh in each direction only had minimal impact on the energies ($\approx\SI{3}{\milli\electronvolt\per\atom}$) and forces ($<\SI{0.1}{\milli\electronvolt\per\angstrom}$), comparable to the error incurred by the NNFF. 
The lattice constants of both cubic phases cover \SIrange{-2}{5}{\percent} of their respective equilibrium values. For the tetragonal phase, deformations in $a$ and $c$ span \SIrange{-1}{4}{\percent} and \SIrange{-1.5}{3}{\percent}, respectively. The monoclinic lattice parameters vary individually in the range of \SIrange{-5}{5}{\percent}.

\section{Results} \label{sec:sectionIII}
\subsection{Neural network validation and transferability}

\begin{figure}
    \centering
    \includegraphics[width=.49\textwidth]{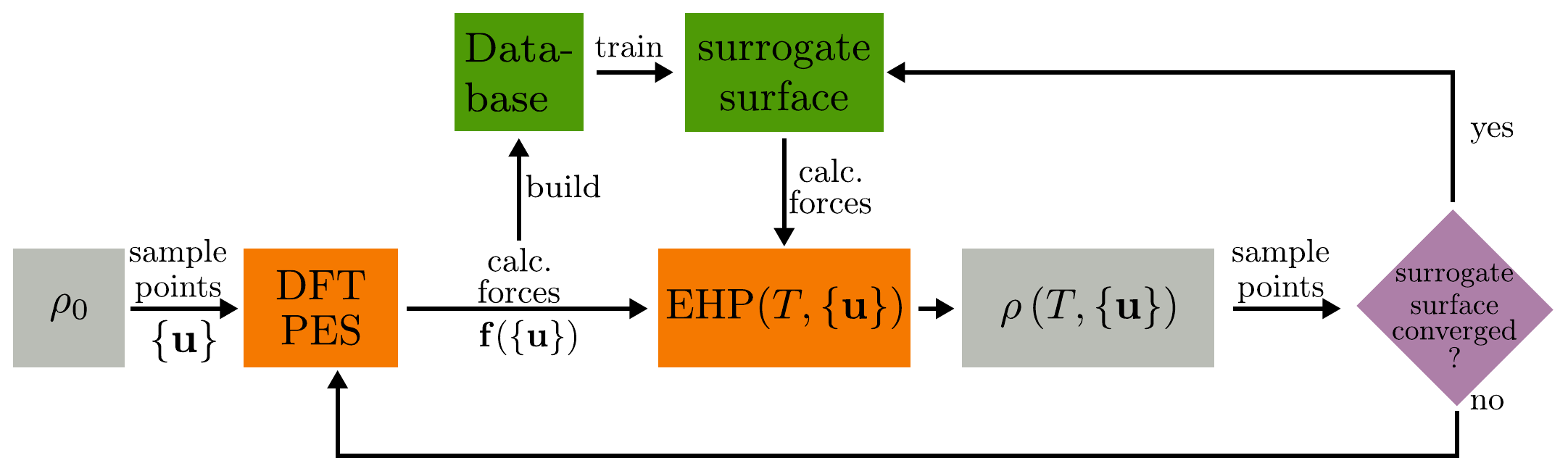}
    \caption{Flowchart of the EHP sampling procedure and gradual construction of a data base used to parametrize the surrogate NNFF.}
    \label{fig:surrogate}
\end{figure}

\begin{figure}
  \includegraphics[width=\columnwidth]{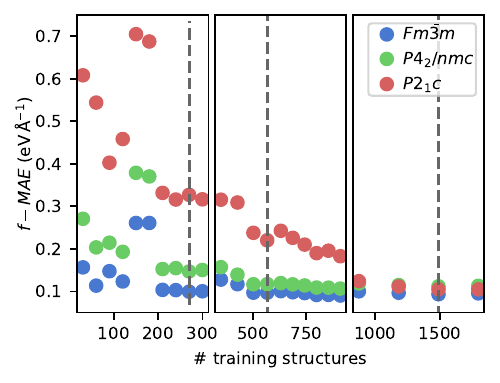}
  \caption{Decrease of force and energy MAE for increasing number of structures. The first panel shows convergence of the error as cubic data is added; the vertical dashed line indicates the chosen number of cubic training structures. In panels two and three this is repeated for tetragonal and monoclinic structures, respectively.}
  \label{fig:cascading}
\end{figure}

To construct a NNFF as a surrogate model for the EHPs we started with a subset, the $T=\SI{2500}{\kelvin}$ samples, of the DFT data generated for our investigation of cI-\ce{HfO_2} from Ref.~\onlinecite{bichelmaier_22}.  As this data was generated $30$ points at a time through an iterative DFT-based EHP scheme, it has a naturally implied hierarchy. Starting out from a density matrix ($\rho_0$) constructed from the FD force constants, we move along this EHP hierarchy as indicated in \cref{fig:surrogate}, adding data points to the electronic structure database after each iteration. Based on this database an NNFF is parameterized at each iteration. The performance is assesed on a test set consisting of \num{50} randomly chosen and unseen structures from each of the phases, $Fm\bar{3}m$, $P4_2nmc$ and $P2_1c$ resulting in the first panel of \cref{fig:cascading}. This process continues, mimicking active learning, until the error has converged for the cubic phase, as is indicated by the bottom arrow in \cref{fig:surrogate}. After convergence, the number of cubic structures contained in the database is fixed at $270$ structures (the dashed line in the first panel of \cref{fig:cascading}). 
Now, using the NNFF the force evaluations for treatment of the cubic phase with the EHP algorithm can be performed, following the top-right arrow in the flowchart (\cref{fig:surrogate}).

Next, we start from this database and repeat the procedure for each $(a,c)$ pair defined for the tetragonal phase, adding $\approx 75$ tetragonal structures sampled at \SI{2500}{\kelvin} to the database at each iteration. In the middle panel of \cref{fig:cascading} it can already be seen that the local environments in the tetragonal and cubic phase have a high degree of similarity, requiring only an additional $\approx 300$ structures until the prediction error stops improving. Finally the same procedure is repeated for all the $(a,b,c)$ combinations of the monoclinic phase. A major improvement in performance for the monoclinic test set can be seen already after the first addition of monoclinic data to the training set (corresponding to the first set of points in the third panel of \cref{fig:cascading}). For convergence a total of $921$ additional monoclinic data points is required.

\begin{figure}
\centering
  \includegraphics[width=\columnwidth]{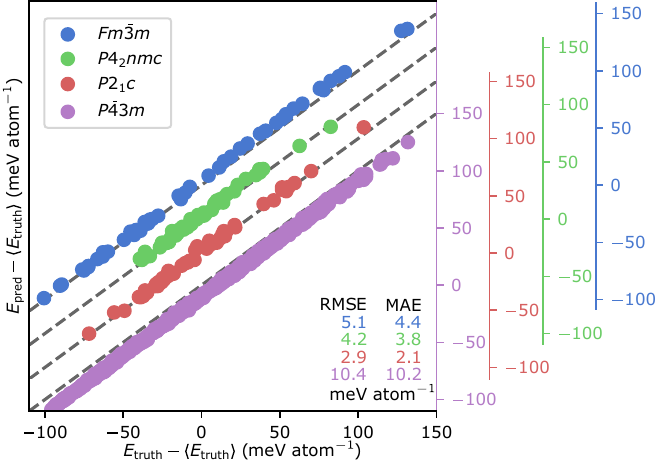}
  \includegraphics[width=\columnwidth]{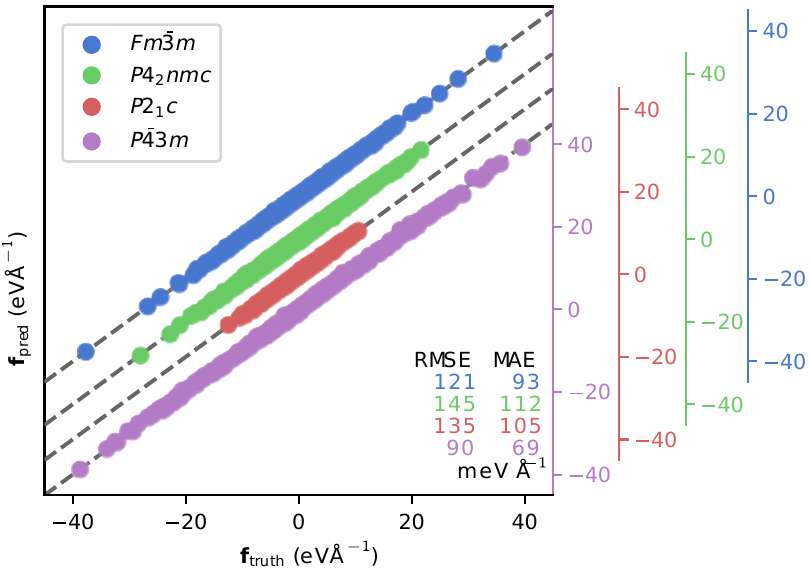}
  \caption{Parity plots and errors of the test set performance for the energies (top) and forces (bottom) for the $Fm\bar{3}m$, $P\bar{4}3m$, $P2_1c$ and $P4_2nmc$ phases of \ce{HfO2}.}
  \label{fig:parity}
\end{figure}

\begin{table}[!htbp]
\centering

\begin{tabular}{*7c}
\toprule
\textbf{Phase} &  \multicolumn{3}{c}{\textbf{Force} (\SI{}{\milli\electronvolt\angstrom^{-1}})} & \multicolumn{3}{c}{\textbf{Energy} (\SI{}{\milli\electronvolt\,\atom^{-1}})} \\
\midrule
{}   & RMSE   & MAE & $\sigma$   & RMSE   & MAE & $\sigma$ \\
$Fm\bar{3}m$   &  174 & 108   & 2989  & 4 & 3  & 185\\
$P4_2/nmc$   &  201 & 141   & 4474  & 5 & 4 & 227\\
$P2_1c$   &  155  &  120   & 2327  &  3 & 2 & 44\\
\bottomrule
\end{tabular}
\caption{Comparison of the root-mean-square error (RMSE), the mean-absolute error (MAE) and the standard deviation ($\sigma$) of the data for the three phases contained in the validation set.}
\label{tab:stats}
\end{table}
This cascading algorithm results in a training set with \num{1487} structures and a validation set with \num{599} structures. The corresponding validation statistics and the standard deviation of the dataset are shown in \cref{tab:stats}.
The parity plots and statistics of the final model applied to said test set, as well as to an additional set of data points for $P\bar{4}3m$, are shown in \cref{fig:parity}. Clearly, the model performs well on the three phases used to generate configurations constituting the training data set, but it also provides excellent predictions for the completely unknown cII phase. This indicates the local environments sampled throughout the training provide enough diversity to result in a transferable model. As an additional qualitative validation metric we show the excellent agreement of the NNFF and DFT-backed finite displacement (FD) and EHP phonon spectrum of the $Fm\bar{3}m$ phase in the supplemental materials \cite{suppl}. Notably, none of the small-displacement FD configurations were part of the training, yet their prediction is sufficient to reconstruct a phonon spectrum.

\subsection{Influence of pressure}
\begin{figure}
    \centering
    \includegraphics[width=\columnwidth]{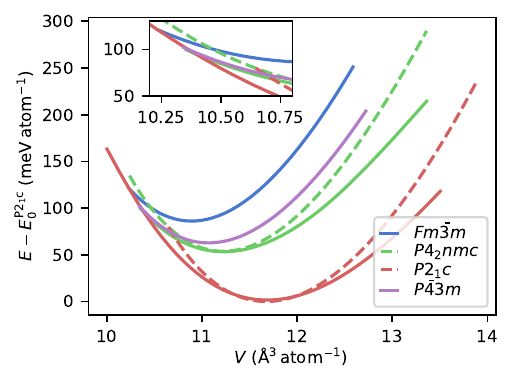}
    \caption{The $(E,V)$-curves of the various phases of \ce{HfO2} under consideration. The solid lines show the actual $(E,V)$ curves, i.e., taking variations of the lattice constants into account, while the dashed lines are only volume-scaled. }
    \label{fig:phases_e_v}
\end{figure}

For a first look at the stability of the four phases considered, energy-volume curves with all atoms at their equilibrium positions are shown in \cref{fig:phases_e_v}. For the tetragonal ($P4_2nmc$) and monoclinic phase ($P2_1c$) we have calculated curves both by fully relaxing all lattice constants at each volume (full lines) and by simply scaling the volumes and keeping the aspect ratios fixed at the optimal volume values (dashed lines). For the $P2_1c$ phase, the effect of the unit cell can be split into the impact of changing the cell lengths and of changing $\beta$. Within the studied volume region \SIrange{10.5}{12.5}{\angstrom^3\,\atom^{-1}}, the impact of $\beta$ relaxation is negligible; hence, $\beta$ is kept constant at its optimal equilibrium angle $\beta_0=\SI{99.69}{\degree}$. Energy-volume curves in literature (e.g. Fig.~4 in \cite{wu_prb_21}) often suggest a first-order phase transition between the $P2_1c$, $P4_2nmc$ and $Fm\bar{3}m$ induced by a pressure of $\approx\SI{8.7}{\giga\pascal}$ when the effect of vibrations is neglected. This, however, only holds if the lattice constants are volume-scaled. With variable cell lengths, the tetragonal and monoclinic phases are favorable over the $Fm\bar{3}m$ until their axes are compressed to the point that they become essentially cubic at \SI{10.3}{\angstrom^3\,\atom^{-1}}. \cref{fig:phases_e_v} also shows that the lower-symmetry cubic $P\bar{4}3m$ phase is lower in energy (by \SI{23}{\milli\electronvolt\,\atom^{-1}}) than $Fm\bar{3}m$ at ambient pressure. Again, no first-order pressure induced phase transition to $Fm\bar{3}m$ is found and the two phases gradually become energetically indistinguishable at volumes around \SI{10.3}{\angstrom^3\,\atom^{-1}}. 

The optimal PBE volumes are given in Table~\ref{tab:lattice}. Extrapolating the experimental data of the monoclinic and tetragonal phase by Haggerty et al. \cite{haggerty_jacs_14} to \SI{0}{\kelvin}, we arrive at equilibrium volumes of $V_0^{\mathrm{exp}}=\SI{11.39}{\angstrom^3\,\atom^{-1}}$ and $V_0^{\mathrm{exp}}=\SI{10.91}{\angstrom^3\,\atom^{-1}}$ for the monoclinic phase and tetragonal phases respectively. The PBE results in Table~\ref{tab:lattice} thus overestimate the volume by $\approx2.5-3.0$\%. The tendency to overestimate unit cell volumes can substantially influence the calculated thermodynamic properties \cite{Wu_PRB04,vanderbilt_94_prl,xie_2022_arxiv} and is a well-known shortcoming of the PBE functional \cite{Wu_PRB04,tran_16}. It has led to the development of more specialized functionals like the Wu-Cohen \cite{WC} and PBEsol \cite{PBEsol} functionals where a lower exchange enhancement leads to volumes in better agreement with experiment. In Table~\ref{tab:lattice} we show that PBEsol underestimates the volumes by approximately 1~\%. 
The most straightforward strategy, however, is the addition of a small artificial isotropic pressure, $p_a$ \cite{vanderbilt_94_prl,Wu_PRB04,xie_2022_arxiv}. Here we choose to perform the calculations with PBE and fix the artificial pressure $p_a=\SI{4}{\giga\pascal}$, resulting in the volumes shown in \cref{tab:lattice}. The thus obtained bulk modulus for the m-phase, $B_0=\SI{184}{\giga\pascal}$ is in agreement with experimental results ranging from \SIrange{187}{197}{\giga\pascal} \cite{bulk1}.
\begin{table}
\begin{tabular}{l|c|c|c|}

 & \multicolumn{2}{c|}{\SI{0}{\giga\pascal}}  & \SI{4}{\giga\pascal}  \\ 
  & \textbf{PBE} & \textbf{PBEsol}  &  \textbf{PBE}  \\ \hline
$Fm\bar{3}m$  & 10.89 &  10.54 &  10.73    \\ \hline
$P\bar{4}3m$   & 11.07   & 10.68  & 10.89 \\ \hline
$P4_2nmc$ &   11.23 & 10.79 & 11  \\ \hline
$P2_1c$   &   11.68 & 11.28 &  11.44    \\ \hline
{} & \multicolumn{3}{c}{$V$ (\SI{}{\angstrom^3\atom^{-1}})}

\end{tabular}
\caption{Volumes of the phases as obtained using PBE and PBEsol functionals at \SI{0}{\giga\pascal} in comparison to those obtained by application of an artificial pressure $p_a=\SI{4}{\giga\pascal}$}
\label{tab:lattice}
\end{table}

\subsection{Influence of temperature}
Besides pressure, temperature is an important contributor to phase stability. Conceptually, the EHP treats the influence of temperature in a manner similar to the quasi-harmonic approach, by calculating phonons on a grid of lattice constants, but with the important difference that a self-consistent phonon spectrum must be calculated for each set of lattice constants and temperatures. Furthermore the free energy contains the correction term accounting for the difference between the ensemble-averaged potential energy and its effective harmonic approximation.

For the investigation of the cubic phase in Ref.~\onlinecite{bichelmaier_22} eight volume points were considered. Using an effective sampling technique to reuse data points, approximately $1200$ single-point calculations were required to achieve convergence for a rather narrow temperature range. Achieving convergence for the full temperature range for a low symmetry structure like the $P2_1c$ can become computationally prohibitive. Using a conservative estimate, the total number of calculations required for all four phases at the current temperature range would exceed $80000$. While possible, this would have been a major undertaking and the advantage of using a comparatively cheap surrogate model becomes apparent. 
The NNFF evaluates a configuration in negligible time and there is no need to be frugal when it comes to the number of samples drawn at each iteration. Consider for example, the $Fm\bar{3}m$ phase, where we evaluated \num{150000} structures. For the tetragonal phase approximately \num{500000} structures were evaluated, and for the monoclinic phase that number was almost \num{2.3} million.
This allows obtaining good effective sample sizes and, with them, a much more accurate evaluation of particularly $F_{\mathrm{corr}}$.

\begin{figure}
    \centering
    \includegraphics[width=\columnwidth]{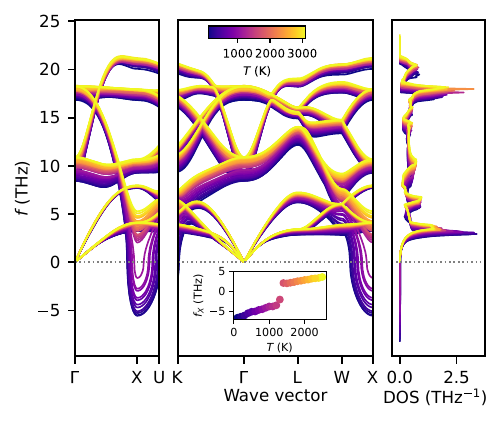}
    \caption{Continuous hardening with temperature of the soft mode exhibited by the $Fm\bar{3}m$ phase at $V=\SI{10.87}{\angstrom^3\atom^{-1}}$. The inset shows the value of the lowest-lying acoustic mode at $X$ as a function of temperature.}
    \label{fig:hardening}
\end{figure}
\begin{figure}
    \centering
    \includegraphics{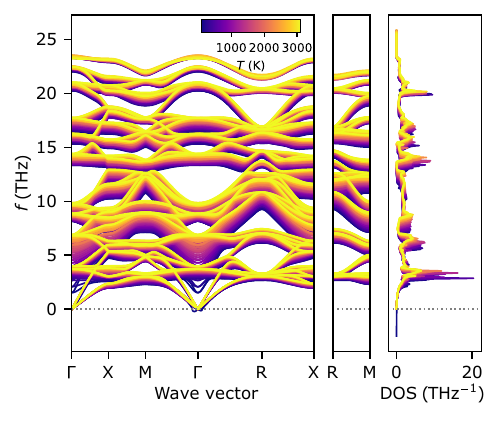}
    \caption{Continuous hardening with temperature of the soft mode exhibited by the $P\bar{4}3m$ phase at $V=\SI{11}{\angstrom^3\atom^{-1}}$.}
    \label{fig:hardening2}
\end{figure}

The finite-displacement phonon band structure of $Fm\bar{3}m$ exhibits imaginary frequencies around the $X$ point \cite{huan_PRB_2014,kuenneth_pss,bichelmaier_22}. This instability at $X=(\frac{1}{2}, \frac{1}{2},\frac{1}{2})$ was linked to a cubic-to-tetragonal deformation in zirconia which is structurally similar \cite{kuwabara_PRB_2005}. This would indicate that the phase is mechanically unstable and would make the application of \cref{eqn:harmonic:free} invalid. We have already shown that the EHP makes $Fm\bar{3}m$ become mechanically stable at temperatures of \SI{2100}{\kelvin} and above \cite{bichelmaier_22}. Using the NNFF as the backend we can employ a much finer temperature mesh and provide a clearer picture. As shown in \cref{fig:hardening}, the lowest $X$-point frequency increases continuously with temperature until it abruptly changes sign at around \SI{1200}{\kelvin}.
Moreover, we find a small instability in the phonon band structure close to the $\Gamma$ point, at $(\frac{1}{30}, \frac{1}{30}, 0)$ and permutations thereof, in the $P\bar{4}3m$ phase when using the finite displacement method (\cref{fig:hardening2}). Clearly, this phase is also mechanically stabilized through temperature, and using the EHP $P\bar{4}3m$ becomes stable at temperatures lower than \SI{1000}{\kelvin}.  

\begin{figure}
    \centering
    \includegraphics[width=\columnwidth]{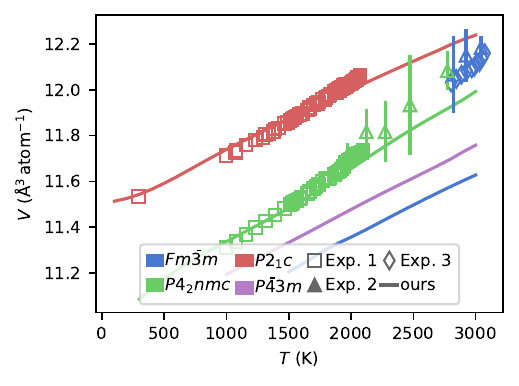}
    \caption{Thermal expansion of the two cubic, the tetragonal and the monoclinic phase obtained using the NNFF as compared to Haggerty et al. (Exp. 1, \cite{haggerty_jacs_14}), Tobase et al. (Exp. 2, \cite{tobase_pss_18}) and Hong et al. (Exp. 3, \cite{hong_nature_2018}).}
    \label{fig:texpansion}
\end{figure}

Having achieved real phonon spectra, the free energies of the phases can be calculated using \cref{eqn:harmonic:free} and the correction term \cref{eqn:correction}. We use the EHP approach for all phases, including those that are mechanically stable in the ordinary harmonic approximation (namely $P2_1c$ and $P4_2nmc$), and 
obtain free energy-volume curves as a function of temperature and phase and hence the thermal expansion. 
As shown in \cref{fig:texpansion}, under the artificial pressure $p_a=\SI{4}{\giga\pascal}$, the thermal expansion of the monoclinic and tetragonal phases very closely matches the results measured by Haggerty et al. \cite{haggerty_jacs_14} and Tobase et al. \cite{tobase_pss_18}. The monoclinic result at room temperature, $V=\SI{11.54}{\angstrom^3\per\atom}$, is further corroborated by measurements performed by Ruh et al. ($V=\SI{11.52}{\angstrom^3\per\atom}$, \cite{ruh}) and Akahama et al. ($V=\SI{11.53}{\angstrom^3\per\atom}$, \cite{bulk1}). Also individual lattice parameters agree well with experiment, as shown in the supplementary information \cite{suppl}. 

Considering the good agreement found for the tetragonal and monoclinic phases, it is somewhat surprising that we see a distinct mismatch with experiment for the presumed cubic \ce{HfO2} phases \cite{tobase_pss_18,hong_nature_2018}. 
Nonetheless, we can draw some important conclusions: First, the rarely studied $P\bar{4}3m$, while still differing by \SI{3}{\percent}, is closer in volume to those measurements than the $Fm\bar{3}m$ phase. Secondly, the phases measured in \cite{hong_nature_2018,tobase_pss_18} might not correspond to stoichiometric \ce{HfO2}. Previous investigations indicate a stabilization of high-symmetry phases over the monoclinic phase particularly with decreasing oxygen content \cite{mittmann_19,kaiser_22,rauwel_12,islamov_19}. Specifically a low-temperature cubic phase of \ce{HfO_{1.7}} is reported in Ref.~\onlinecite{kaiser_22}. Furthermore, oxygen-deficient, but stoichiometrically unspecified, nanoparticles of cubic hafnia have been obtained through synthesis in a reductive solvent \cite{rauwel_12}. These oxygen-deficient structures could in turn be affected by vacancy-mediated chemical expansion, as discussed extensively in literature \cite{marrocchelli_pccp_12,dilpuneet_cms_15,aschauer_prb_13}, which would explain the discrepancy visible in \cref{fig:texpansion}. 

\subsection{Thermodynamic phase stability}
\begin{figure}
    \centering
    \includegraphics[width=\columnwidth]{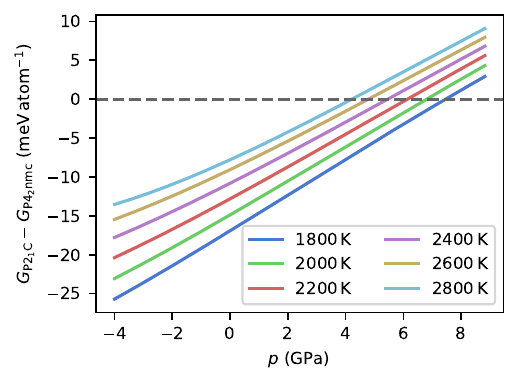}
    \caption{The differences in free enthalpy of the monoclinic and tetragonal phase of \ce{HfO2}. The x-axis includes an artificial pressure of $p_a=\SI{4}{\giga\pascal}$ as discussed in the text.}
    \label{fig:p21c_v_p42nmc}
\end{figure}

Using the free energy-volumes curves we furthermore find that temperature stabilizes the tetragonal $P4_2nmc$ phase compared to the $P2_1c$ phase, in agreement with experiment \cite{massalski1986binary,wang_jamcersoc_06}. However, as seen in \cref{fig:p21c_v_p42nmc}, the application of additional pressure is essential to stabilize the $P4_2nmc$ phase over the $P2_1c$ within the temperature window studied in this work. For example, at \SI{2600}{\kelvin}, a pressure of roughly \SI{4}{\giga\pascal} in addition to $p_a$ is needed to stabilize the tetragonal phase. This can again potentially be attributed to the chosen functional, PBE, which tends to produce slightly enlarged cells. A further indication supporting this is that the pressure needed to take the PBE equilibrium volume of the $Fm\bar{3}m$ phase to the PBEsol equilibrium volume, is of the same order of magnitude (\SI{7.7}{\giga\pascal}). 

\begin{figure}
    \centering
    \includegraphics[width=\columnwidth]{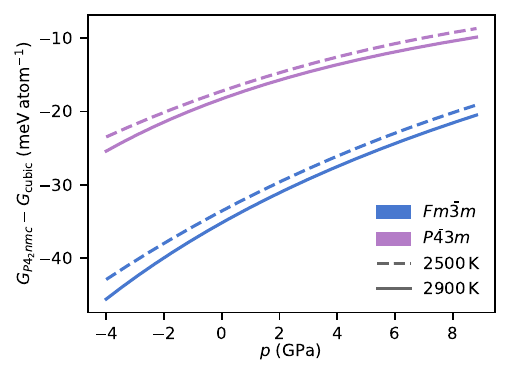}
    \caption{Impact of temperature, \SI{2500}{\kelvin} (dashed) and \SI{2900}{\kelvin} (solid), and pressure on the free enthalpy difference between $Fm\bar{3}m$ (blue), $P\bar{4}3m$ (violet) and $P4_2nmc$.}
    \label{fig:cubic_v_tet}
\end{figure}

Interestingly, we do not find that temperature stabilizes the two studied cubic phases, $Fm\bar{3}m$ and $P\bar{4}3m$, compared to the tetragonal phase, \cref{fig:cubic_v_tet}.
While the EHP methodology, due to the harmonic ansatz used in the construction of the density matrix, is not able to capture all anharmonic contributions, the effect of further corrections would have to be very significant to overcome the energy differences required for a transition to a cubic lattice. Not even the studied pressure ranges are able to stabilize the cubic lattices, as is shown in \cref{fig:cubic_v_tet}. 
On the other hand, the energy differences are small in comparison with the characteristic scale of thermal fluctuations per degree of freedom (e.g. the $P4_2nmc$-to-$P\bar{4}3m$ enthalpy difference at \SI{4}{\giga\pascal} is approximately \SI{12}{\milli\electronvolt\per\atom}, but $k_B T$ at \SI{2000}{\kelvin} is $\approx\SI{172}{\milli\electronvolt\per\atom}$). Processing-induced oxygen vacancies, or, at temperatures this high, even temperature-induced oxygen vacancies \cite{hong_nature_2018,kumar_jpd_22,Hadacek_2007} might stabilize the cubic phase over its tetragonal counterpart, in particular in combination with interface and surface strain of the fine powders typically used in XRD measurements.
Future research should more closely investigate how the energy balance between $Fm\bar{3}m$ and $P\bar{4}3m$ shifts when those parameters are considered. Likewise, the impact potential contributions to this picture arising from configurational entropy in the presence of defects should be probed. As a simple estimate $TS_{\mathrm{conf}}=Tk_B\left[\left(1-c\right)\log{\left(1-c\right)}+c\log c\right] \approx \SI{28}{\milli\electronvolt\per\atom}$ at an assumed defect concentration of $c=\SI{10}{\percent}$.

These results do not preclude metastability as might be achieved in experiment: We emphasize that for a phase to be stable within the EHP formalism, its free energy needs to be the lowest. This might even be the case in molecular dynamics under certain conditions, such as in the work by Fan et al. \cite{fan_2019} using ab-initio molecular dynamics (AIMD), in which they claim to observe a transition to $Fm\bar{3}m$. However, the computational restrictions inherent to AIMD might limit the interpretability of these results. First of all, the use of a small simulation box precludes relaxation through long-wavelength deformations. 
Secondly, the molecular-dynamics trajectory quickly scans through temperatures or pressures without a guarantee that the system is allowed to fully relax.
To the extent a cubic phase is observed in Ref.~\onlinecite{fan_2019}, the methodology is furthermore insufficient to conclusively establish whether it corresponds to $Fm\bar{3}m$ or $P\bar{4}3m$. Those limitations, imposed by the high computational cost of direct ab-initio molecular dynamics, highlight the value of an accelerated surrogate model.

\section{Conclusion}{\label{sec:conclusion}}
We have described a framework to construct neural network force fields (NNFF) using effective harmonic potentials (EHP) and applied it towards the parametrization of a transferable NNFF for \ce{HfO2}. The potential performs comparably to state-of-the-art, with an out-of-sample force-MAE of $\approx$$\,\SI{100}{\milli\electronvolt\per\angstrom}$ and an energy-MAE of $\approx$$\,\SI{10}{\milli\electronvolt\per\atom}$, despite being trained on only $\approx$$\,1500$ structures. The NNFF is then applied in conjunction with EHPs to study phase stability of the ambient phases of \ce{HfO2}. The methodology leads to agreement with experimental literature regarding the thermal expansion of the $P2_1c$ and $P4_2nmc$ phases. At roughly \SI{4}{\giga\pascal}, which is comparable to an effective pressure linking PBE and PBEsol, we find a monoclinic-to-tetragonal phase transition at a temperature comparable to experiment (\SI{2500}{\kelvin} vs. \SI{2050}{\kelvin}). We furthermore conclude that the most likely stoichiometric high-temperature space group of cubic bulk \ce{HfO2} is $P\bar{4}3m$, as opposed to the traditionally assumed $Fm\bar{3}m$. Nonetheless, we do not see a tetragonal-to-cubic phase transition within the temperature and pressure range studied in this work and find that the observed lattice constants of the cubic phases are substantially larger than what is calculated for the corresponding stoichiometric phases. Future studies should explore the potential of metastability (e.g. through molecular dynamics) and the impact of oxygen vacancies on the various phases. 

\section*{Acknowledgements} \label{sec:acknowledgements}
AI4DI receives funding within the Electronic Components and Systems for European Leadership Joint Undertaking (ESCEL JU) in collaboration with the European Union’s Horizon2020 Framework Programme and National Authorities, under grant agreement n° 826060. This work was furthermore supported by the Austrian Science Fund (FWF) (SFB F81 TACO).

%

\end{document}